\documentclass[twocolumn,superscriptaddress,showpacs,amsmath,amssymb,aps,prl,floatfix,prstab,prstper,nolongbibliography]{revtex4-2}

\usepackage{color} 
\usepackage{graphicx}
\usepackage{braket} 
\usepackage{dcolumn}
\usepackage{bm}
\usepackage{subfigure}
\usepackage{amssymb}
\usepackage{multirow}
\usepackage{natbib}
\usepackage{amsmath}
\graphicspath{{plots/}}
\usepackage[english]{babel}

\usepackage[colorlinks,linkcolor=blue,anchorcolor=blue,citecolor=blue,urlcolor=blue,CJKbookmarks=True]{hyperref}

\begin{document}
	
	\title{Full-scale \textit{ab initio} simulations of laser-driven atomistic dynamics}
	
	\author{Qiyu Zeng} 
	\author{Bo Chen}
	\author{Shen Zhang}
	\author{Dongdong Kang}
	\affiliation{College of Science, National University of Defense Technology, Changsha, Hunan 410073, China}
	\affiliation{Hunan Key Laboratory of Extreme Matter and Applications, National University of Defense Technology, Changsha 410073, China}
	\author{Han Wang}
	\affiliation{Laboratory of Computational Physics, Institute of Applied Physics and Computational Mathematics, Beijing 100088, China}
	\author{Xiaoxiang Yu} \email{xxyu@nudt.edu.cn} 
	\author{Jiayu Dai} \email{jydai@nudt.edu.cn} 
	\affiliation{College of Science, National University of Defense Technology, Changsha, Hunan 410073, China}
	\affiliation{Hunan Key Laboratory of Extreme Matter and Applications, National University of Defense Technology, Changsha 410073, China}
	
	\date{\today}
	
	\begin{abstract}
		
		The coupling of excited states and ionic dynamics is the basic and challenging point for the materials response at extreme conditions. In laboratory, the intense laser produces transient nature and complexity with highly nonequilibrium states, making it extremely difficult and interesting for both experimental measurements and theoretical methods. With the inclusion of laser-excited states, we extended \textit{ab initio} method into the direct simulations of whole laser-driven microscopic dynamics from solid to liquid. We constructed the framework of combining the electron-temperature-dependent deep neural network potential energy surface with hybrid atomistic-continuum approach, controlling non-adiabatic energy exchange and atomistic dynamics, which enables consistent interpretation of experimental data. By large scale \textit{ab inito} simulations, we demonstrate that the nonthermal effects introduced by hot electrons play a dominant role in modulating the lattice dynamics, thermodynamic pathway, and structural transformation. We highlight that the present work provides a path to realistic computational studies of laser-driven processes, thus bridging the gap between experiments and simulations.
		
	\end{abstract}
	\maketitle

\section{Introduction}

Intense laser-matter interaction plays an important role in many applications including inertial confinement fusion \cite{shawareb2022}, laser micromachining \cite{gattass2008}, and material synthesis \cite{guan2022}. Ultrafast laser excitation can drive matter into extremely non-equilibrium states, in which the hot electron and cold lattice coexist. The subsequent atomistic dynamics is therefore a long-standing challenge, because it is governed by the interplay between excited-electron-modulated potential energy surface (PES) \cite{recoules2006}, electron-ion coupling \cite{liu2022}, and geometric characteristics of irradiated samples \cite{ivanov2003effect}. 

Tremendous efforts based on time-resolved probing techniques and simulations have provided valuable insights into the nonthermal behaviors \cite{ernstorfer2009, giret2014, zhang2016ultrafast, chen2021ultrafast, ono2021}, kinetics of laser-driven melting \cite{mo2018heterogeneous, mo2019, wu2022,arefev2022}, and electron-phonon coupling \cite{cho2016, smirnov2020, molina2022}. The related processes from cold solid to hot liquid and plasma are the typical multiscale dynamics due to the cascade of interrelated processes triggered by the laser excitation, both in time scale and size scale. Therefore, it is of great difficulty and importance to construct a well-coordinated picture between experimental and theoretical efforts. For example, the dynamics of laser-excited Au is still under debates \cite{recoules2006, ernstorfer2009, daraszewicz2013}, regarding the phonon behaviors driven by laser excitation. In these cases, different priori assumptions on material response were  usually made \cite{medvedev2020, mahfoud2021}. 

The above obscure stems from the technical limitations that the present methods can not capture both the nonthermal and intrinsic scale of laser-induced process at the same time. For \textit{ab initio} methods such as time-dependent density functional theory (DFT), the sizes are limited to $10^1\sim10^3$ atoms and $10^2\sim10^4\ \rm{fs}$, unable to access realistic representation of structural transformations of irradiated samples. While for classical molecular dynamics simulations coupled with two-temperature model (TTM-MD) \cite{ivanov2003combined, zeng2020structural}, the implementation of empirical potential like embedded-atom-method (EAM) is limited in prior knowledge and model complexity, thus can hardly capture the high dimensional dependence of PES on both atomic local environments and electron occupations for a wide range of temperature and density \cite{daraszewicz2013, murphy2015}, leading to the inadequate description of nonthermal nature of laser-driven processes. Therefore, bringing the advantage of \textit{ab initio} and large-scale molecular dynamics including nonthermal effect becomes the route one must take. 

\begin{figure*}[htbp]
	\centering
	\includegraphics[width=0.9\linewidth]{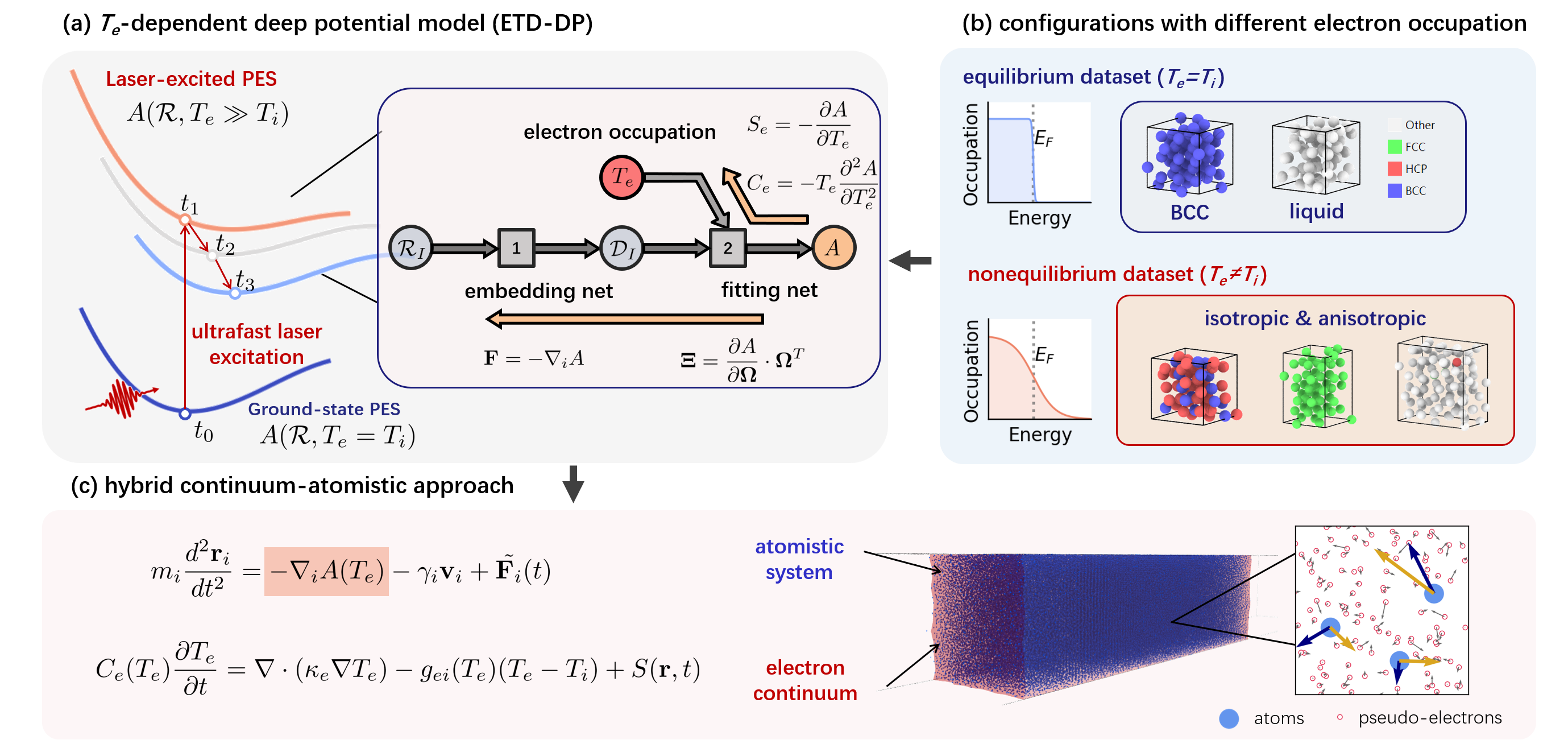}
	\caption{ \textbf{Schematic diagram of workflow for efficient and accurate simulation of laser-driven atomistic dynamics.} (a) ETD-DP model. $T_e$ is the electron temperature, regarding to the electron occupation distribution. The free energy $A$, force $\mathbf F$, virial $\Xi$, electronic entropy $S_e$, and electronic heat capacity $C_e$ can be inferred through backpropagation algorithm. (b) iterative concurrent learning scheme is used to efficiently sample atomic configurations for a wide range of equilibrium and non-equilibrium conditions. (c) hybrid atomistic continuum approach. The evolution of electron subsystem allows atomistic system transits between different PES, and the Langevin thermostat is introduced to mimic non-adiabatic energy exchange between electron and lattice.}
	\label{fig:1}
\end{figure*}

In this paper, we developed \textit{ab initio} atomistic-continuum model by combining two-temperature-model (TTM) with an extended deep potential molecular dynamics (DPMD), as illustrated in Fig.\ref{fig:1}. When the ultrafast laser interacts with solids, the electrons quickly thermalized at timescaes of femtoseconds, producing highly non-equilibrium states (electron temperature $T_e$ $\gg$ ion temperature $T_i$). The hot $T_e$ will result in the redistributed charge density firstly and then modifies the PES of ions. To capture this physics, we introduced the laser-excited PES by constructing electron-temperature-dependent deep neural network potential firstly, and coupled the PES into additional electron continuum subsystem via TTM-MD framework. By this way, we can directly simulate the whole electron-ion coupled dynamics during the laser-driven processes with large-scale simulations within \textit{ab initio} accuracy. We take tungsten as an example and systematically validate the accuracy of our model in describing lattice dynamics, thermophysical properties, and laser heating process in both equilibrium and laser-excited states, by comparing with the related experimental results recently \cite{mo2019}. 

\section{Results}

\textbf{Construction of laser-excited PES.} 
Recent efforts have demonstrated the success of machine learning model towards large-scale simulations of \textit{ab initio} quality at extreme conditions \cite{zeng2021ab, chen2021atomistic, yang2022, chen2022three,redmer2022,Cheng2023,hinz2023,plettenberg2023}, but most of the studies focus on the equilibrium-state and ground-state applications. Here the electron-temperature-dependent deep potential (ETD-DP) model is implemented in the framework of deep potential method \cite{zhang2018deep, zhang2020warm, zeng2021ab} to model the laser-driven dynamics. 

To avoid constructing hand-crafted features or kernels for different types of bulk systems, a general end-to-end symmetry preserving scheme is adopted \cite{zhang2018end}. As illustrated in Fig.\ref{fig:1}(a), the ETD-DP model consists of an embedding network and a fitting network. The embedding network is designed to transform the coordinate matrices $\mathcal R_I$ to symmetry preserving features, encoded in the descriptor $\mathcal D_I$. And the fitting network is a standard fully connected feedforward neural network, mapping the descriptor to the atomic contribution of total energies.

The newly introduced parameter, electron temperature $T_e$, is used to characterize the laser modulation on PES, in which electron occupation distribution is far away from electron-ion equilibrium states. This ETD-DP is defined as 
\begin{align}
	A = A(\mathcal R, T_e) = \sum_i \mathcal N_{\alpha_i} (\mathcal D_{\alpha_i}(r_i,\{r_j\}_{j\in n(i)}), T_e)
\end{align}
where $A(\mathcal R, T_e)$ is the potential energy depends on the local atomic environment ($\mathcal R$) and $T_e$, $\mathcal N_{\alpha_i}$ denotes the neural network of specified chemical species of $\alpha_i$ of atom $i$, and the descriptors $\mathcal D_{\alpha_i}$ describes the symmetry preserved local environment of atom $i$ with its neighbor list $n(i)=\{j|r_{ji}<r_{cut}\}$, respectively. 

To generate an ETD-DP, the new degree of freedom, $T_e$, will dramatically expand the sampling space in the data labelling process, introducing expensive computational costs. Therefore, an iterative concurrent learning scheme \cite{zhang2019active} is highly required to efficiently sample atomic configurations under both equilibrium ($T_e=T_i$) and non-equilibrium conditions ($T_e \ne T_i$). As shown in Fig.\ref{fig:1}(b), to explore the density-temperature space with different electron occupations $(\rho,T_i,T_e)$, a variety of crystal structures are used as the initial configurations to run multiple DPMD simulations. And an ensemble of ETD-DP is trained with the same dataset but with different parameter initializations. The model deviation, denoted as the maximum standard deviation of the predicted atomic forces by the ensemble of ETD-DP, is used to evaluate whether the explored atomic configurations should be send to generate referenced \textit{ab initio} energies, forces, and virial tensors.

\textbf{Two-temperature model coupled DPMD (TTM-DPMD).} To model the whole ultrafast laser-driven processes from cold solid to plasma, we should couple quantum electron subsystem and strongly coupled ionic subsystem. Here, we implemented our laser-excited PES into the TTM-MD framework \cite{ivanov2003combined, duffy2007, zeng2020structural}, going beyond traditional ground-state EAM and neural-network-driven PES descriptions. As shown in Fig.\ref{fig:1}(c), the heat conduction equation of electron continuum characterizes the temporal evolution of electron occupations, thus governing the transition of ionic system between different $T_e$-dependent PES. Langevin dynamics is incorporated to mimic the dynamic electron-ion collisions \cite{duffy2007, dai2010unified, dai2012dynamic, zeng2020structural}. The TTM-DPMD is defined as follows, 
\begin{align}
	C_e(T_e)\frac{\partial T_e}{\partial t} &= \nabla \cdot(\kappa_e \nabla T_e) - g_{ei}(T_e) (T_e-T_i) + S(\mathbf r, t) \\
	m_i\frac{d^2 \mathbf r_i}{dt^2} &=-\nabla_i A(T_e) -\gamma_i \mathbf v_i + \tilde {\mathbf F_i}(t)
\end{align}
where $C_e$ is the electronic heat capacity, $\kappa_e$ the electronic thermal conductivity, $g_{ei}$ the electron-phonon coupling constant, $S(\mathbf r, t)$ the laser source. The ions evolves on the $T_e$-dependent PES $A(\mathcal R, T_e)$, and suffers fluctuation-dissipation forces $-\gamma_i\mathbf v_i + \tilde {\mathbf F_i}(t)$ from electron sea. Here $\gamma_i$ is the friction parameter that characterizes the electron-ion equilibration rate, relating to the electron-phonon coupling constant through $\gamma = g_{ei} m_i/2n_i k_B$, where $n_i$ the ion number density. The $\tilde {\mathbf F_i}(t)$ term is a stochastic force term with a Gaussian distribution, whose mean and variance is given by
$\langle \tilde {\mathbf F_i(t)} \rangle = 0$ and $\langle \tilde {\mathbf F_i}(t) \cdot \tilde {\mathbf F_i}(t') \rangle = 2\gamma_i k_B T_e \delta (t-t')$.

In TTM-DPMD, by practically choosing the electron temperature or ionic temperature in the meshgrid as the additional parameter in ETD-DP model, the ions can evolve under laser-excited PES ($T_e\gg T_i$) or ground-state PES ($T_e=T_i$), so that we can separate the nonthermal effects defined by the electronic excitation from thermally driven atomic dynamics and phase transformation.

\textbf{Validating neural network model for laser-excited tungsten.}
To validate the effectiveness of extended DP model, we chose tungsten as our target system. Tungsten is a typical transition metal, with half-filled \textit{d} bands that is sensitive to $T_e$. Upon laser excitation, tungsten is expected to go through a complicated dynamic process including possible nonthermal solid-solid phase transition \cite{giret2014, murphy2015, medvedev2020, mahfoud2021}, attracting much attention but remains ambiguous. Here we generate a $T_e$-dependent deep-neural-network tungsten model by learning from DFT data calculated with the generalized gradient approximation (GGA) of the exchange-correlation functional \cite{pbeGGA} using VASP package \cite{vasp}. The atomic configurations used in the training set are collected from a wide range of $(\rho, T_i, T_e)$ condition, covering the phase space of the body-centered-cubic (BCC), close-packed structure, uniaxially distorted crystalline, and the liquid structures. More details about DP training can be found in the supplemental materials \cite{supple}.

\begin{figure}[tb]
	\centering
	\includegraphics[width=0.8\linewidth]{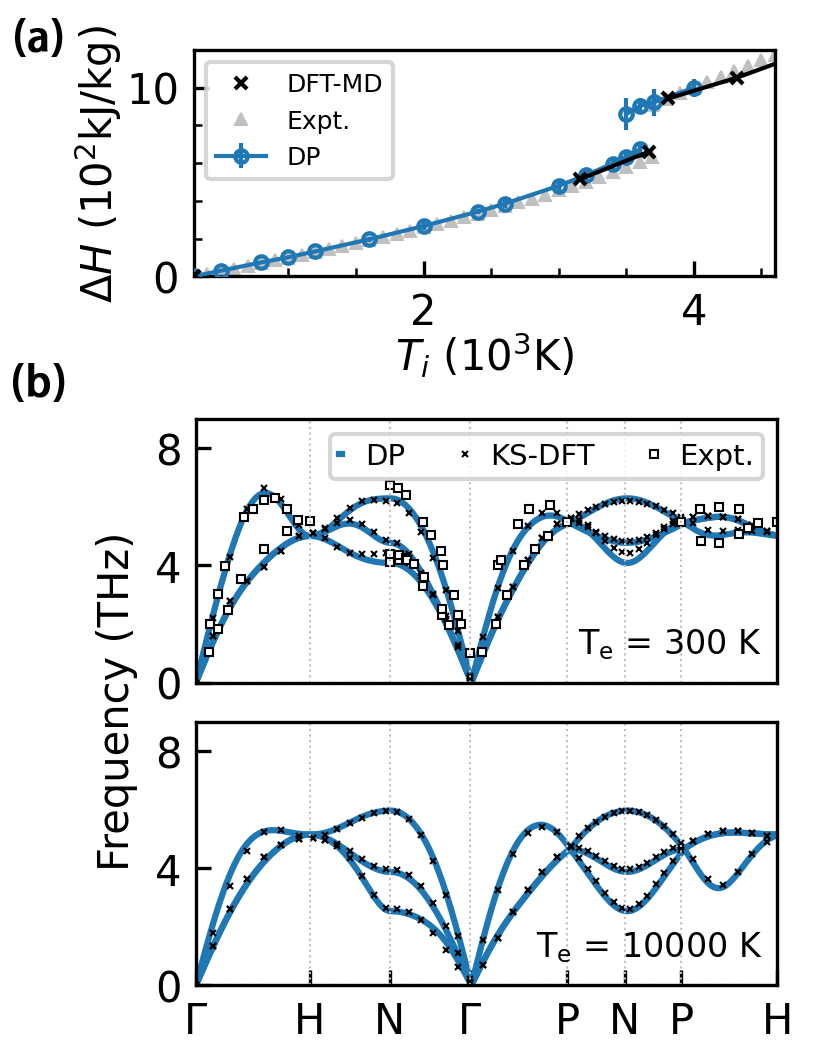}
	\caption{ \textbf{Validating accuracy of ETD-DP model.} (a) Temperature dependence of enthalpy under isobaric heating ($p=1\ \rm{bar}$) with the reference temperature of 300 K. The blue line, the black cross, and grey square denotes the DPMD results, previous DFT-MD prediction \cite{minakov2018}, and isobaric expansion experimental data \cite{arblaster2018} respectively. (b) phonon dispersion of laser-excited tungsten ($\rho_0=19.15\ \rm{g\ cm^{-3}}$). The black cross and white squares represent the individual KS-DFT calculation and experimental measurements \cite{larose1976}. }
	\label{fig:2}
\end{figure}

Here we pay special attention to thermodynamic properties of equilibrium tungsten that are closely related to laser heating process. The melting temperature predicted by ground-state DPMD (3550 K \cite{supple}) is in consistence to the previous DFT-MD (3450 $\pm$ 100 K \cite{wang2011melting}) and Gaussian approximation potentials simulations (3540 K \cite{byggmastar2020}), which confirms that the present PES can reproduce melting with DFT accuracy. Furthermore, the dependence of DPMD-predicted enthalpy on temperature along isobaric heating condition is shown in Fig.\ref{fig:2}(a), and the experimental data agree very well with our DPMD predictions, especially in the liquid regime \cite{arblaster2018}. The estimated enthalpy of fusion at the melting point ($\Delta H_m=237\pm20\ \rm{kJ/kg}$) is also close to the DFT-MD values ($250\ \rm{kJ/kg}$) \cite{minakov2018} and other experimental values (see Table.S1 in \cite{supple}).

Based on calculated thermophysical properties, we can determine the complete melting threshold $\epsilon_m$, which is the laser energy that is sufficient to drive the complete melt of the samples. We found $\epsilon_m = 0.92\pm0.04\ \rm{MJ/kg}$, corresponding absorbed pump fluence is $53.0\pm2.2\ \rm{mJ/cm^2}$ for 30-nm-thick tungsten film \cite{supple}. Such values are in agreement with the estimated values from experimental results \cite{mo2019, white1984, berthault1986,hixson1990,kaschnitz1990,arblaster2018}, in which energy density is approximately $0.94\ \rm{MJ/kg}$ (pump fluence of $53.8\ \rm{mJ/cm^2}$). The density decrease at elevated temperature as shown in Fig.S3 \cite{supple}, is also consistent to the experimental measurements \cite{hixson1990, hupf2008, kaschnitz1990}.  

The lattice dynamics, that requires high-order derivatives of PES, were further investigated. As shown in Fig.\ref{fig:2}(b), the phonon dispersion curves of BCC tungsten under both equilibrium and non-equilibrium states are well-reproduced compared with the DFT results. In particular, comparing with the phonon dispersion at $T_e = 300\ \rm{K}$, the directional phonon softening is observed along the $\rm{H-N}$ and $\rm{N-\Gamma}$ path in the first Brillouin zone at elevated electron temperature ($T_e = 10000\ \rm{K}$), which can be attributed to the delocalization half-filled $d$ bands \cite{giret2014, murphy2015}. The depopulation of such a strong directional component in electronic bonding weakens the directional forces and may drive the crystalline structures towards to close-packed forms. These results indicate that the neural network PES can provide faithful prediction related properties in consistency to experiments or \textit{ab initio} method.

\begin{figure}[tb]
	\centering
	\includegraphics[width=0.8\linewidth]{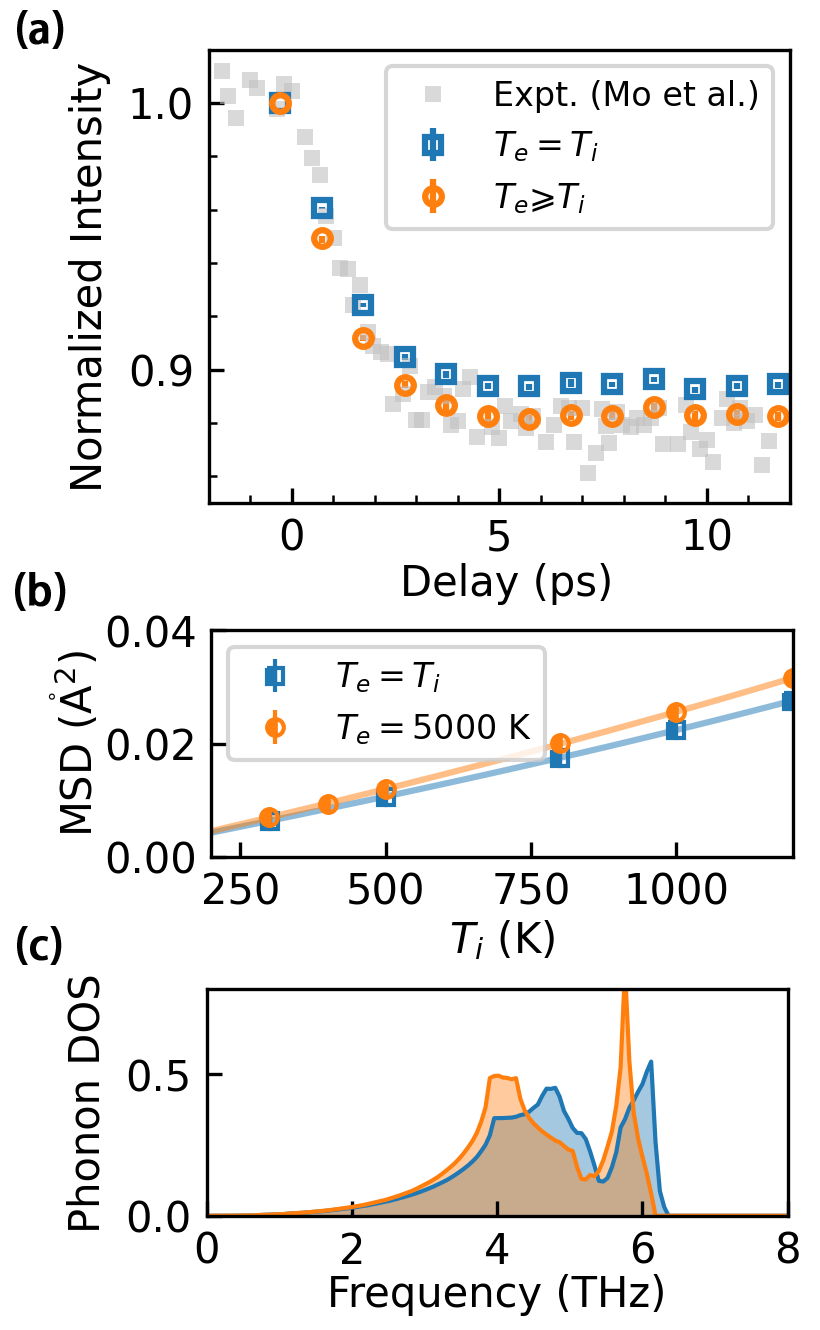}
	\caption{ \textbf{Capturing nonthermal effect with TTM-DPMD approach.} Comparison of (a) temporal evolution of (211) diffraction peak intensity in structure factor under absorbed laser energy density of $0.08\ \rm{MJ/kg}$, compared with experimental data \cite{mo2019}. (b) Temperature dependence of mean square displacement with isobaric constrains and (c) phonon density of states (PDOS), obtained under equilibrium condition (blue) and nonequilibrium condition (orange).}
	\label{fig:3}
\end{figure}

\textbf{Direct \textit{ab initio} simulations of laser-driven dynamics.}	It is stressed that our explicit electron-temperature-dependence PES can well capture the nonthermal nature of laser-excited metals. When implemented in TTM-DPMD framework, it allows us to establish a comprehensive understanding on laser-induced non-equilibrium states within \textit{ab initio} accuracy. Recent time-resolved ultrafast electron and X-ray diffraction experiments collects direct quantitative structural information of laser-driven processes \cite{mo2019}, providing a benchmark for the validation of the present newly-developed methods. Here, we apply TTM-DPMD to directly simulate the dynamic response of tungsten nanofilm under different absorbed laser energy densities.

\begin{figure*}[htbp]
	\centering\includegraphics[width=0.9\linewidth]{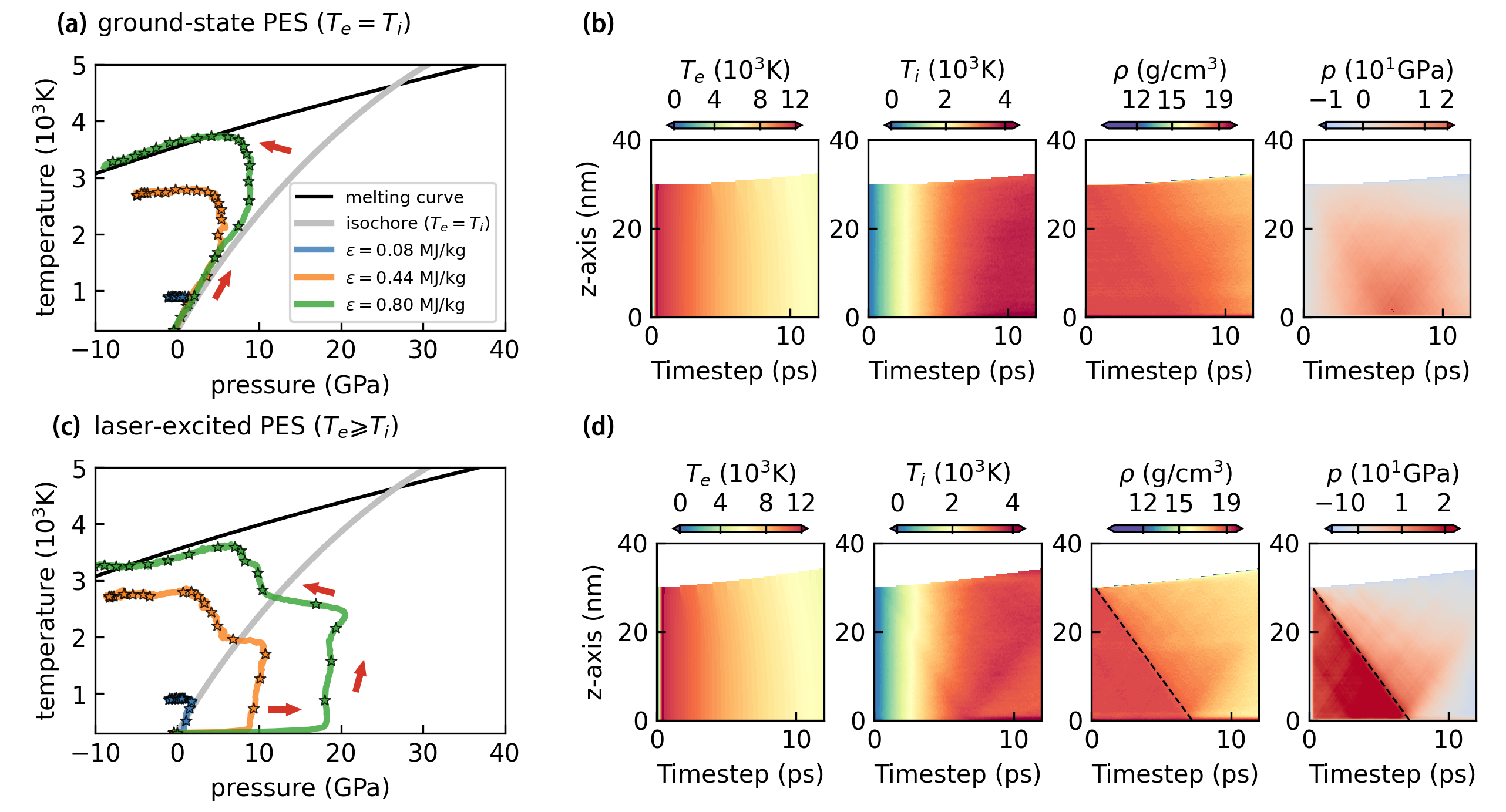}
	\caption{ \textbf{Hot electron modifies the thermodynamic pathway.} Comparison of (a)(c) thermodynamic pathway (b)(d) temporal evolution of thermodynamic profile of nanofilm, predicted by ground-state PES and laser-excited PES, respectively. In (a)(c), the red arrows indicate the evolution path of selected regime (z = 14.0 nm) in the tungsten nanofilm, and the thermodynamic state is highlighted by colored stars every 1 ps. In (d), the black dashed lines are used to highlight the propagation of stress waves, whose slope represents a constant propagation speed of $\sim 4.3\ \rm{km/s}$.} 
	\label{fig:4}
\end{figure*}

In TTM-DPMD simulations, full-scale \textit{ab initio} description in one-dimension of polycrystalline (PC) 30-nm-thick tungsten nanofilm is considered, according to relevant UED experiment \cite{mo2019}. For PC systems, large size included 752,650 atoms is used to describe crystal grains with random shapes, orientations, and different types of boundaries. The size of each grain ranges from $\sim 5\ \rm{nm}$ to $\sim 7\ \rm{nm}$ and each grain contains more than $10^4$ atoms (totally reaching to millions of atoms), which cannot be achieved by the traditional time-dependent DFT simulations. Moreover, extra 30 nm vacuum space perpendicular to laser incident direction is set to allow free surface response to the internal stress relaxation, and extra spring forces are introduced for atoms in the bottom regime relating to their initial lattice site, to present bonding to the substrate (see supplementary materials \cite{supple}). 

Considering the ballistic transportation of excited electron in tungsten (the mean free path $\sim 33\ \rm{nm}$), we assumed the uniform deposition of laser energy with relatively low energy density of $0.08\ \rm{MJ/kg}$ (corresponding to absorbed laser fluence of $4.8\ \rm{mJ/cm^2}$). In this case, a moderate two-temperature state is created at the initial stage, where maximum electron temperature can reach to 4400 K. Through electron-ion energy exchange, the system quickly reaches thermal equilibrium ($T_e=T_i\sim920\ \rm{K}$) at t = 5 ps.

The structure factor is calculated to extract the decay dynamics of Laue diffraction peak (LDP) \cite{coleman2013, supple}, which is an important quantity to diagnose the structural dynamics in experiments \cite{mo2019}. As shown in Fig.\ref{fig:3}(a), based on TTM-DPMD simulations with the inclusion of laser-driven excited states, the temporal evolution of normalized intensity of (211) LDP agrees well with UED measurements. Conversely, the results from simulations by ground-state PES deviate from experimentally measured values significantly. It is interesting to say that the thermal process ($T_e=T_i$) exhibits remarkably slower decay dynamics than the process with excited states ($T_e\ge T_i$) upon such laser fluence. By further investigating the lattice vibration of bulk tungsten, we note that even under moderate non-equilibrium state ($T_e=5000\ \rm{K}$), a relative increase of over 10\% in mean square displacement (MSD) can be observed under isobaric heating condition, as shown in Fig.\ref{fig:3}(b). The enhancement of lattice vibration can be attributed to the hot-electron-induced phonon softening (Fig.\ref{fig:3}(c)). Such nonequilibrium and nonthermal effects therefore modify the dynamics of diffraction signals according to Debye-Waller formula \cite{supple}, in which the decay of LDP is relating to temporal evolution of lattice temperature and temperature dependence of MSD.  The quantitative consistency between our simulations and experiments validates our model further, and then provide a chance to further elucidate laser excitation effects.

By increasing laser energy density up to $0.80\ \rm{MJ/kg}$, the irradiated tungsten nanofilm starts with more severe nonequilibrium states ($T_e=11200\ \rm{K},T_i=300\ \rm{K}$). As presented in Fig.\ref{fig:4}(a)(b), the evolution of tungsten nanofilm predicted by ground-state PES ($T_e=T_i$) is a purely thermal process governed by electron-ion coupling. With increased lattice temperature, the system firstly evolves along the equilibrium isochore in the first 4 ps, where the ionic kinetic pressure accumulates to $\sim 10\ \rm{GPa}$. Then the thermal pressure is gradually released due to existence of free surface. Although the thermal expansion process leads to temperature and density decrease, the gradient in thermodynamic profile is slight and the whole system can be considered as homogeneous.

When laser-induced changes in the PES is included ($T_e\ge T_i$), the thermodynamic pathway and thermodynamic profile is totally different. As shown in the Fig.\ref{fig:4}(c)(d), the ultrafast excitation of electrons results in the buildup of extra pressure on a sub-picosecond timescale (more details in Fig.S7). Such hot-electron-contributed pressure increases monotonically with increased laser energy density, from $\sim1\ \rm{GPa}$ with $T_{e,0} = 4400\ \rm{K}$ ($\epsilon=0.08\ \rm{MJ/kg}$) to $\sim17\ \rm{GPa}$ with $T_{e,0} = 11200\ \rm{K}$ ($\epsilon=0.80\ \rm{MJ/kg}$). The tungsten nanofilm then quickly responds to this nonthermal internal stress, triggering anisotropic volume relaxation dynamics. As a result, a significant inhomogeneity is demonstrated in the thermodynamic profiles. In Fig.\ref{fig:4}(d), the propagation and reflection of stress waves can be identified with velocity of $\sim 4.3\ \rm{km/s}$, accompanied with the density decrease of $\sim 1\ \rm{g/cm^3}$. With existence of free surface, the build up and uniaxial relaxation of nonthermal stress can strongly influence the thermodynamic pathway especially under high laser fluence, which cannot simply be assumed to be isochoric or isotropically isobaric. We highlight that such real-time material response captured by TTM-DPMD simulation provides unique insights into previous controversial issues on nonthermal behavior of laser-excited matter \cite{daraszewicz2013, medvedev2020}.

\section{Discussion}

In this work, we developed the deep learning model to perform large-scale \textit{ab inito} simulations on the laser-induced atomistic dynamics, with quantum accuracy on the non-thermal effects. To validate the accuracy, special attention is paid to recent experiments. We successfully reproduce the experimental data with our model. It is therefore verified that the laser-excited states have profound effects on the thermodynamic evolution and structural transformation dynamics. More importantly, the combination of deep learning techniques with hybrid continuum-atomistic approach bridges the theoretical method and experimental observations, providing a new path to establish accurate and complete understanding of the atomistic dynamics under ultrafast laser interactions.

\section{Method}

\textbf{DP training.}
The ETD-DP models for tungsten are generated with DeePMD-kit packages \cite{wang2018deepmd} by considering $T_e$ as atomic parameter. Deep Potential Generator (DP-GEN) \cite{zhang2019active}, has been adopted to sample the most compact and adequate data set that guarantees the uniform accuracy of ETD-DP in the explored configuration space. 
We consider BCC structure (54 atoms) and liquid structure (54 atoms) as the initial configurations and run DPMD under NVT and NPT ensemble (both isotropic and uniaxial constrains are considered), where temperatures ranges from 100 K to 6000 K, pressure ranges from -15 to 60 GPa, and corresponding electronic temperature ranges from 100 K to 25000 K. The training sets consist of 6366 configurations under equilibrium condition ($T_e = T_i$) and 6820 configurations sampled under two-temperature state ($T_e > T_i$).

For DP training, the embedding network is composed of three layers (25, 50, and 100 nodes) while the fitting network has three hidden layers with 240 nodes in each layer. The total number of training steps is set to 400 000. The radius cutoff $r_c$ is chosen to be $6.0\ {\rm \mathring A}$. The weight parameters in loss function for energies $p_e$, forces $p_f$, and virials $p_V$ are set to $(0.02, 1000, 0.02)$ at the beginning of training and gradually change to $(1.0, 1.0, 1.0)$.

The self-consistency calculations are all performed with the VASP package \cite{giannozzi2017advanced}. The Perdew-Burke-Erzerhof (PBE) exchange correlation functional is used \cite{perdew1996generalized}, and the pseudopotential takes the projector augmented-wave (PAW) formalism \cite{blochl1994projector, holzwarth2001projector}. The sampling of Brillouin zone is chosen as 0.2 ${\rm \mathring A^{-1}}$ under ambient conditions ($T\le 300\ \rm{K}$), and 0.5 ${\rm \mathring A^{-1}}$ for high temperature. 

\textbf{TTM-DPMD simulation setting.}
We perform TTM-DPMD simulations with LAMMPS package \cite{plimpton1995fast} through modified EXTRA-FIX packages \cite{zeng2020structural}.
The electronic heat capacity is calculated by individual DFT calculations $C_e=T_e\frac{\partial S_e}{\partial T_e}$, which is consistent with previous calculations \cite{lin2008}. The electron-phonon coupling factor is set to constant ($G_0 = 2.0\times10^{17}\ \rm{W\ m^{-2}\ K^{-1}}$) according to relevant ultrafast electron diffraction experiments \cite{mo2019}. 
The electron thermal conductivity is described by the Drude model relationship, $\kappa_e(T_e,T_i)=\frac{1}{3} v_F^2 C_e(T_e) \tau_e(T_e,T_i)$, where $v_F$ is Fermi velocity and $\tau_e(T_e,T_i)$ is the total electron scattering time defined by the electron-electron and electron-phonon scattering rates, $1/\tau_e=1/\tau_{e-e}+1/\tau_{e-ph}=AT_e^2 +BT_i$. The coefficients $A=2.11\times10^{-4}\ \rm{K^{-2}\ ps^{-1}}, B=8.4\times 10^{-2}\ \rm{K^{-1}\ ps^{-1}},v_F=9710\ \rm{\mathring A\ ps^{-1}}$ are adopted \cite{grossi2019}.
The duration of laser pulse is set to 130 fs. Since the mean free path of laser-excited electrons is $\sim 33\ \rm{nm}$ in tungsten \cite{steinhogl2005}, the electrons are heated uniformly due to the ballistic transport. Therefore, optical penetration of laser energy can be neglected for simplicity.

For atomic system, the simulation size of polycrystalline sample is set to $30\ \rm{nm}\times20\ \rm{nm}\times20\ \rm{nm}$, containing 752650 atoms, with extra 30 nm vacuum space along the x direction to minic the free boundary condition. 
Extra spring forces are introduced for atoms in the bottom $5\ \rm{\mathring A}$ relating to their initial lattice site to present bonding to the substrate. 

\textbf{phonon spectra calculation.} 
To validate the accuracy of ETP-DP model, we investigate the lattice dynamics that need high order derivatives of PES. 
We use finite displacement method to calculate the phonon dispersion with ALAMODE package \cite{tadano2014} as a postprocessing code. The forces are calculated in $5\times 5 \times 5$ supercell with cell lattice parameter $a_0 = 3.17104 \rm{\mathring A}$. The atomic displacement is set to 0.01 $\rm{\mathring A}$, and the interatomic force constants are extracted from KS-DFT and DPMD calculation respectively. The dynamical matrices are derived from these force displacement data to obtain phonon dispersion spectra.

\textbf{Ultrafast electron diffraction pattern.}
To extract the decay of Laue diffraction peak (LDP) intensities as in the UED experiments, we performed the ultrafast electron diffraction simulations with DIFFRACTION package \cite{coleman2013} to obtain the structure factor $S(Q_x, Q_y)$ defined as follows,
\begin{align}
S &= \frac{F^*F}{N} \\
F(\mathbf Q) &= \sum_i f_i(\mathbf Q;\lambda) e^{i 2\pi \mathbf Q \cdot \mathbf r_i} 
\end{align}
where $\mathbf Q=(Q_x,Q_y,Q_z)$ the wave vector, $f_i$ the atomic scattering factor, $\lambda$ the wavelength of incident electron, $\mathbf r_i$ the coordinates of atom $i$. Here, simulated 3.2MeV electron radiation ($\rm{\lambda \sim 0.34\ pm}$) is used to create selected area electron diffraction (SAED) patterns according to relevant experiments \cite{mo2019}, and the SAED patterns aligned on the [100] axis ($Q_z =0$) are constructed by selecting reciprocal lattice points intersecting a $0.01\rm \mathring A^{-1}$ thick Ewald sphere slice.Detailed discussion can be found in Fig.S5 and Fig.S6 in SI.

\bibliography{refs} 

\section{Acknowledgments}
This work was supported by the National Natural Science Foundation of China under Grant Nos. 11874424, 11904401, 12104507, 12304307, 12122103, and the Science and Technology Innovation Program of Hunan Province under Grant No. 2021RC4026.

\section{Author contributions}
Q.Z and J. D. proposed the original idea and designed the research. Q. Z. carried out the simulations, Q. Z., X. Y., and J.D. analyzed and interpreted the results. Q. Z, X. Y. and J. D. wrote the manuscript with the help from other authors. H. W. contributed to the development of TTM-DPMD method. B. C., S. Z., and D. K. provided additional support for the interpretation of the results. All the authors reviewed the manuscript.

\section{Competing interests}
The authors declare no competing interests.

\section{Additional information}
\textbf{Supplementary information} is available for this paper

\end{document}